\documentclass[aps,prx,twocolumn,multicol,groupedaddress,superscriptaddress,showpacs]{revtex4-1}

\usepackage{graphicx}
\usepackage{color}
\usepackage{amsmath}
\usepackage{amsfonts}
\usepackage{amssymb}
\usepackage{dcolumn}

\usepackage{booktabs}
\begin{document}

\title{How to suppress undesired synchronization}

  \author{V. H. P. Louzada}
    \email{louzada@ethz.ch}
    \affiliation{Computational Physics, IfB, ETH-Honggerberg, Wolfgang-Pauli-Strasse 27, 8093 Zurich, Switzerland}

  \author{N. A. M. Ara\'ujo}
    \email{nuno@ethz.ch}
    \affiliation{Computational Physics, IfB, ETH-Honggerberg, Wolfgang-Pauli-Strasse 27, 8093 Zurich, Switzerland}

  \author{J. S. Andrade, Jr.}
    \email{soares@fisica.ufc.br}
    \affiliation{Computational Physics, IfB, ETH-Honggerberg, Wolfgang-Pauli-Strasse 27, 8093 Zurich, Switzerland}
    \affiliation{Departamento de F\'isica, Universidade Federal do Cear\'a, 60451-970 Fortaleza, Cear\'a, Brazil}

  \author{H. J. Herrmann}
    \email{hans@ifb.baug.ethz.ch}
    \affiliation{Computational Physics, IfB, ETH-Honggerberg, Wolfgang-Pauli-Strasse 27, 8093 Zurich, Switzerland}
    \affiliation{Departamento de F\'isica, Universidade Federal do Cear\'a, 60451-970 Fortaleza, Cear\'a, Brazil}
 
\date{\today}

\begin{abstract}
It is delightful to observe the emergency of synchronization in the blinking of fireflies to attract partners and preys. Other charming examples of synchronization can also be found in a wide range of phenomena such as, e.g., neurons firing, lasers cascades, chemical reactions, and opinion formation. However, in many situations the formation of a coherent state is not pleasant and should be mitigated. For example, the onset of synchronization can be the root of epileptic seizures, traffic congestion in communication networks, and the collapse of constructions. Here we propose the use of contrarians to suppress undesired synchronization. We perform a comparative study of different strategies, either requiring local or total knowledge of the system, and show that the most efficient one solely requires local information. Our results also reveal that, even when the distribution of neighboring interactions is narrow, significant improvement in mitigation is observed when contrarians sit at the highly connected elements. The same qualitative results are obtained for artificially generated networks as well as two real ones, namely, the Routers of the Internet and a neuronal network.

\end{abstract}


\maketitle

\section{Introduction}
In the year 2000, Londoners were presented with the Millennium Bridge, a futuristic footbridge that became the center of attention on the inauguration day. The elbowing of the crowd, eager to be the first to cross it, forced the synchronization of walkers causing a lateral swing of the structure \cite{Strogatz05}. Once on this wobbly structure, how could one avoid such uncomfortable situation? Abnormal synchronization is also the origin of neurological diseases such as epilepsy and Parkinson \cite{Glass01}. Brain pacemakers have been developed and implanted in the patient to discharge an electrical signal into the brain tissue and restore the normal activity \cite{Kringelbach07,McIntyre04,Volkmann02}. But imagine a, still to develop, device able to interact with individual neurons. What would be the best strategy to break the synchronization? A third source of inspiration can be found in the Internet, where several interconnected Routers receive and redistribute the information packages in the network. When multiple Routers synchronize their delivering events, the network collapses, a dysfunction known as TCP global synchronization. To avoid it, several algorithms have been developed and implemented in some Routers \cite{Floyd93}. What is the fraction of such proactive Routers required to avoid global synchronization? In social context, avoiding synchronization might represent a political tool to fight a charismatic leader. Consider a speech that inflames a crowd. Initially every individual claps at his/her own rhythm but rapidly a coherent clap emerges~\cite{Neda00}. If a set of political adversaries (contrarians) try to destroy the harmony, what would be the best strategy, the proper amount of contrarians, and their spatial distribution in the hall? 

The Kuramoto model has extensively been used as the paradigm to study synchronization \cite{Pikovsky03,Acebron05,Osipov07,Arenas08}. In a first attempt to address the questions raised above, we generalize this well-established model to include contrarians which try to suppress the emergence of global synchronization. We present a systematic study of how the synchronizability depends on the fraction of contrarian oscillators for two different strategies and analyze the influence of the topology in the mitigation process. To illustrate our results, contrarian oscillators have been studied \emph{in silico} for two real networks, namely, the Routers that compose the Internet \cite{Lyon03} and the network of neurons of the organism \emph{C. elegans} \cite{Chen06,Bhatla09}. Our results suggest that local contrarians can be used as a powerful way to control synchronization, avoiding the necessity of monitoring the global state. Moreover, spreading contrarians as hubs is also much more effective.

\section{Model}
The described examples are characterized by a set of $N$ oscillators (walkers, neurons, Routers, or spectators), mutually interacting. Hereafter we take the example of the walkers but the model can be straightforwardly extended to all other cases. The stepping of each walker $i$ is characterized by the phase $\theta_i(t)$ and its natural frequency $\omega_i$, corresponding to the stepping frequency when isolated. When the crowd moves, all walkers initially step at their natural frequency but herding (under strong coupling) rapidly leads to coherent walking \cite{Strogatz05}. In the Kuramoto model, the motion of each oscillator is described by a phasor $e^{i\theta_i(t)}$, where $\theta_i(t)$ is the phase, and the coupling between walkers is such that the dynamics of each is governed by,
\begin{equation}
\dot{\theta}_i=\omega_i+\lambda\sum_{j=1}^{N}A_{ij}\sin\left(\theta_j-\theta_i\right),
\label{eq::kuramoto}
\end{equation}
where the sum is over all other walkers ($i\neq j$), $\lambda$ is the coupling strength, $\omega_i$ is the walker's natural frequency, and $\mathbf{A}$ is the connectivity matrix such that $A_{ij}=1$ if walker $i$ is influenced by walker $j$ or zero otherwise. The collective walking can be characterized by the complex order parameter defined as,
\begin{equation}
r(t) e^{i\Psi(t)}=
\frac{1}{N}\sum_j^N e^{i\theta_j(t)},
\label{eq::order.param}
\end{equation}
where the sum is over all walkers, the amplitude ${0\leq|r(t)|\leq1}$ measures the global coherence, and $\Psi(t)$ is the average phase.

To account for contrarians we introduce a second population of $N_c$ walkers also coupled with the others but following a different dynamics. A contrarian $k$ is also characterized by the phase $\theta_k(t)$ and its natural frequency $\omega_k$. We consider two types of coupling: a mean-field (Model~\textit{A}) and a pairwise (Model~\textit{B}). In the mean-field coupling the dynamics of contrarians is governed by,
\begin{equation}
\dot{\theta}_k=\omega_k+\lambda\sin\left(\Psi-\theta_k-\delta\right),
\label{eq::meanfield}
\end{equation}
where $\Psi(t)$ is the average phase and $\delta$ is a phase shift. In the pairwise coupling the dynamics of contrarians is governed by,
\begin{equation}
\dot{\theta}_k=\omega_k+\lambda\sum_{j=1}^{N_T}A_{kj}\sin\left(\theta_j-\theta_k-\delta\right),
\label{eq::pairwise}
\end{equation}
where the sum is over all the $N_T$ walkers ($N_T=N+N_c$). Hereafter, we take $\delta=\pi$ in both cases. Such phase shift between two walkers ($k$ and $j$) would correspond to a walking such that when $k$ steps with the left foot $j$ steps with the right. It is noteworthy that these two models yield different types of frustration. While in Model~A frustration between regular and contrarian walkers is mediated by the average phase, in Model~B the frustration results from a pairwise interaction between regular and contrarian walkers where the former attempts to mutually synchronize while the latter tries to dephase.

Two models accounting for frustration in the mean-field Kuramoto model have recently been discussed in the context of a mixture of positive and negative couplings. Zannette \cite{Zanette05} considered a pairwise coupling where the strength and sign of the interaction between two oscillators is symmetric. This model, in the limit $\omega_k=0$, is equivalent to a magnetic XY model with a distribution of couplings. Hong and Strogatz proposed a different scheme, where regular walkers are also solely coupled with the average phase $\Psi(t)$ and the spatial distribution of regular walkers is not considered \cite{Hong11,Hong11b}. This model is similar to the mean-field limit of Model A discussed here. In contrast to the model discussed here, synchronization cannot be suppressed in any of these previous models.

\section{Results}
In the absence of contrarians, the classical Kuramoto model is characterized by the emergence of synchronization at a critical coupling $\lambda_c$ which depends on the distribution of natural frequencies ($\omega$) and on the degree. While under weak coupling ($\lambda<\lambda_c$) the motion is incoherent ($r=0$), above the critical coupling a coherent motion emerges ($r>0$). In the limit of very strong coupling ($\lambda\gg\lambda_c$) all oscillators participate in the coherent motion. 

\begin{figure}
\includegraphics[width=\columnwidth]{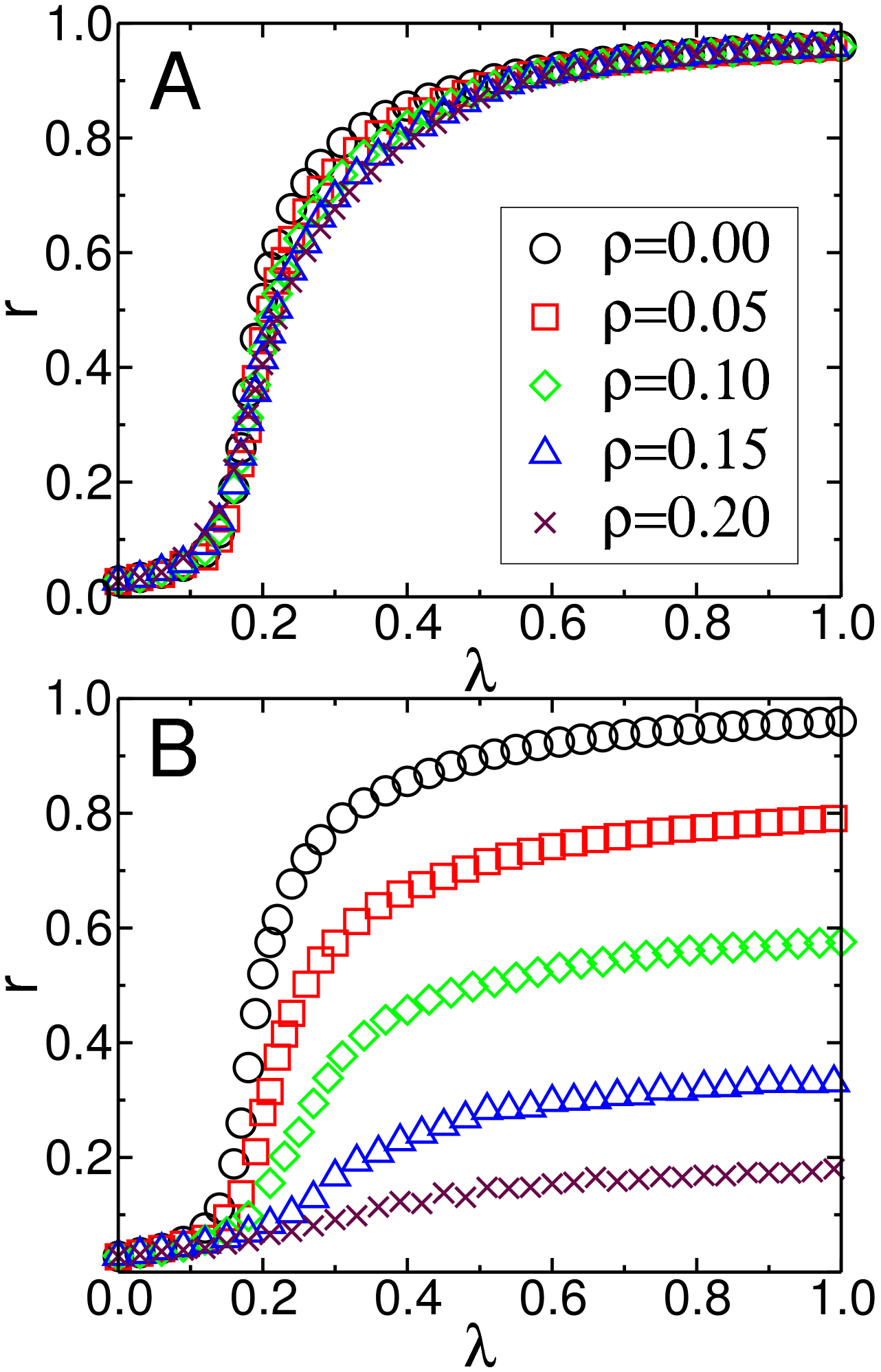}
\caption{\textbf{Comparison between mean-field and pairwise coupling.} Dependence of the order parameter $r$ on the coupling strength $\lambda$, for the mean-field (A) and the pairwise (B) couplings. Different curves stand for different fractions of contrarians $\rho$ randomly distributed in a random graph with average degree equal to four.}
\label{fig::r_x_l_meanfield_pairwise}
\end{figure}

The presence of contrarians can affect the coherent motion. In Figure~\ref{fig::r_x_l_meanfield_pairwise}, different fractions $\rho = N_c/N$ of contrarian oscillators are considered in the mean-field (A) and pairwise (B) models.
While mean-field contrarians are not able to reduce the value of $r$, a fraction as small as $5\%$ of pairwise contrarians is enough to significantly reduce the synchronizability. Further investigation shows that, although contrarians enable the system to desynchronize, mean-field contrarians drive the system to a polarized state, where oscillators are concentrated around two phases: $-\pi$ and $\pi$. It is possible to understand this splitting through the analysis of the stable point (given by $\dot{\theta_k}=0$) for contrarians in the mean-field model, yielding
\begin{equation}
\lambda \sin(\Phi-\theta_k-\pi)=-\omega_k.
\label{eq::dottheta.zero1}
\end{equation}
Assuming that $\omega_k$ is symmetrically distributed around zero, this equation shows that the difference between the phase of contrarians and the average phase must be equal to $\pi$. Thus, contrarians have a tendency towards the extremes of the possible phases, dragging their conformist neighbors in the process. Hence, mean-field contrarians introduce differences in the dynamic behavior of oscillators by polarizing them in two distinct phases (see Supplementary Fig.~S1).

\begin{figure}
\includegraphics[width=\columnwidth]{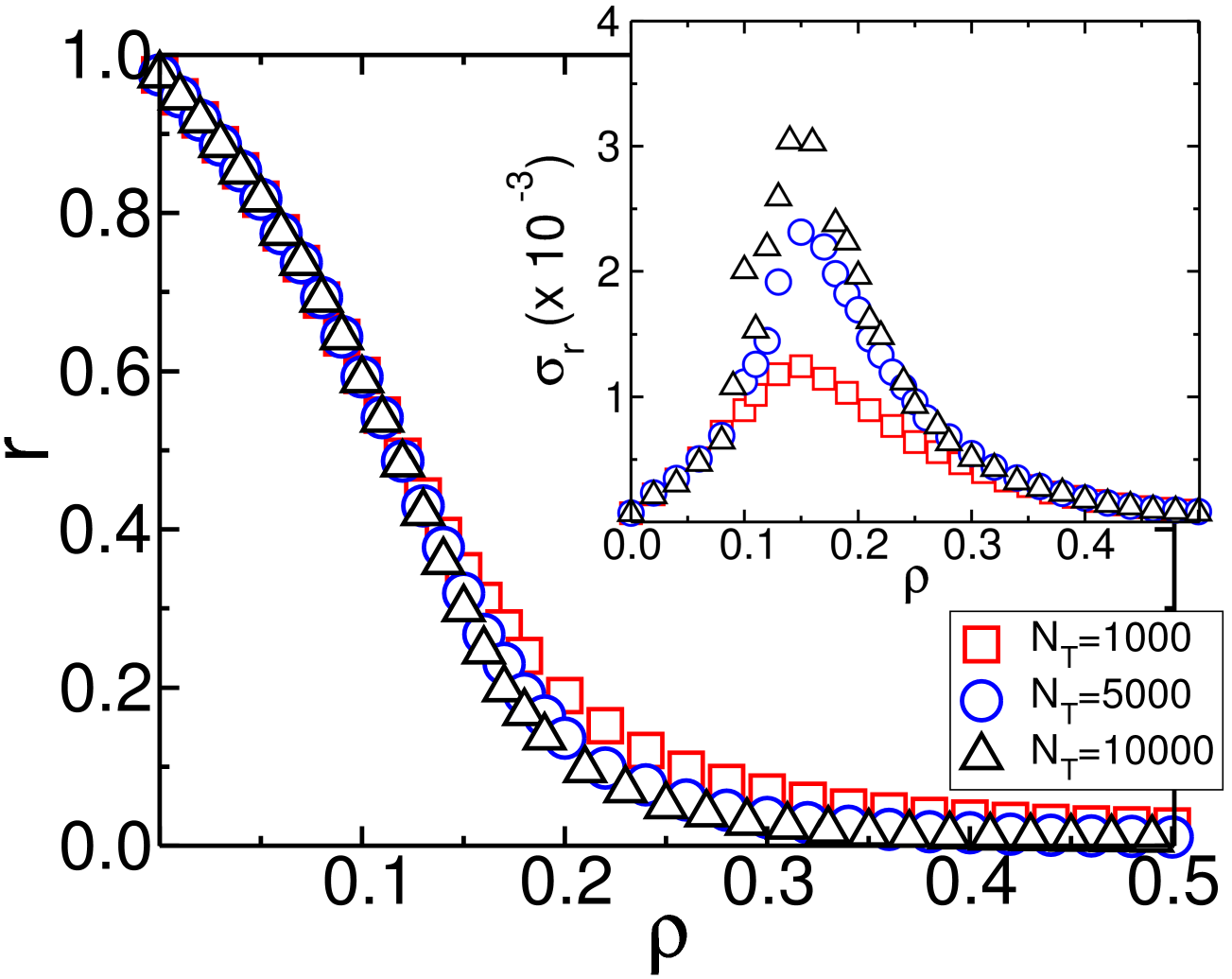}
 \caption{\textbf{Impact of pairwise contrarians on the synchronization.} Order parameter $r$ dependence on the fraction of contrarians $\rho$ showing suppression of synchronization. Different curves stand for different network sizes. The inset contains a plot of the standard deviation of $r$ among samples showing a transition around $\rho=0.15$. A coupling strength of $\lambda=2.0$ has been used.}
\label{fig::r_x_den_221}
\end{figure}
\begin{figure*}
\includegraphics[width=2\columnwidth]{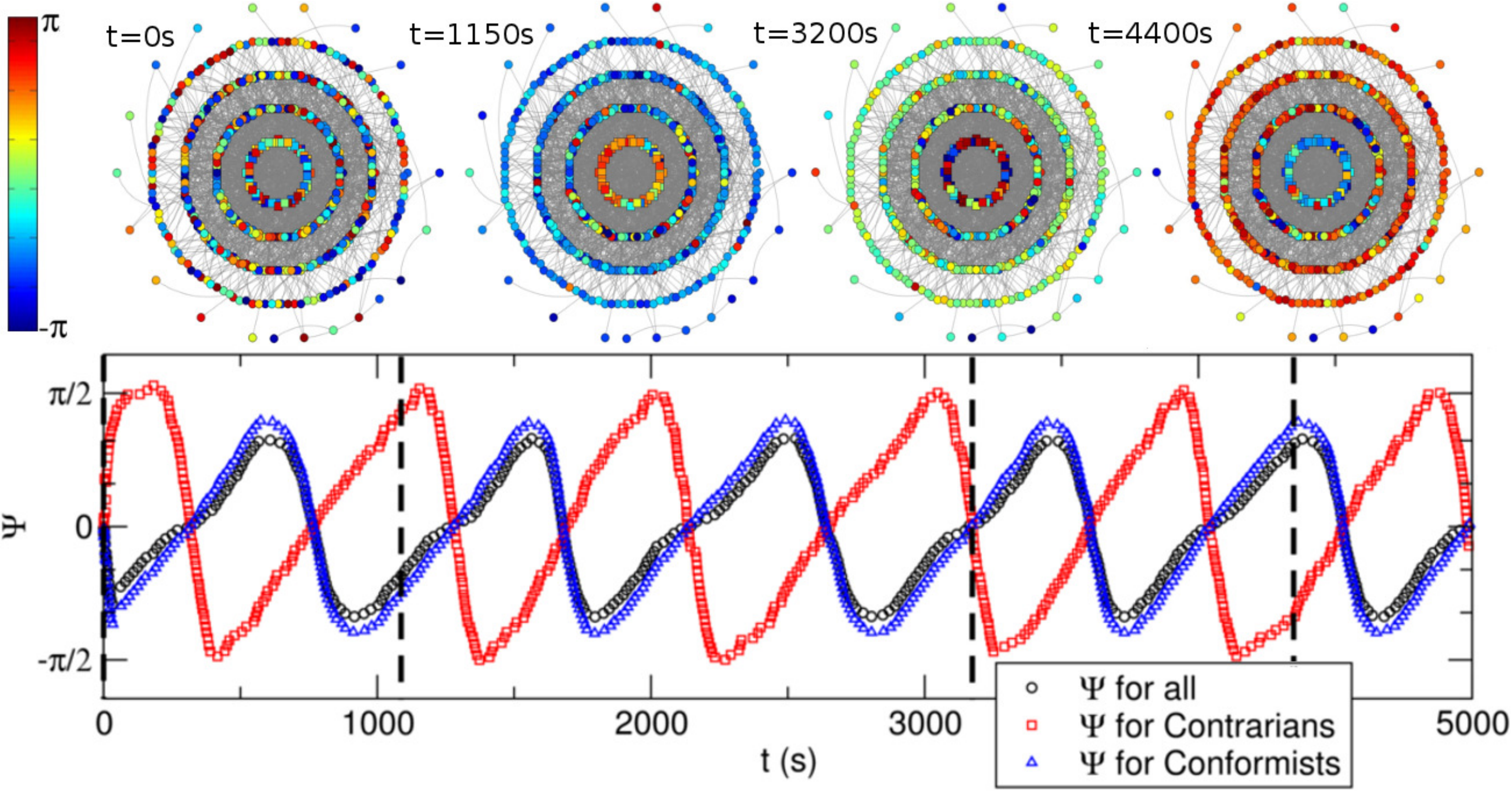}%
\caption{\textbf{Time dependence of the phase.} \textbf{Upper part}: Snapshot of a network of $200$ oscillators at 4 different time steps (vertical dashed lines in the lower part). A total of $20$ pairwise contrarians are displayed in the central layer. Each concentric layer $i$, from inside to outside, contains the $i$th neighbors of the contrarians. The color of each node represents its phase. \textbf{Lower part}: Time dependence of the average phase of contrarians (red squares), conformists (blue triangles), and the whole set of oscillators (black circles) showing a periodic oscillation over time.}
\label{fig::phi_x_t_nets_over_time_pairwise}
\end{figure*}

For the pairwise (B) contrarians, the emergence of a coherent state is suppressed above a certain fraction of contrarians (Figure~\ref{fig::r_x_den_221}). We notice that synchronization is suppressed for values of $\rho>0.15$, as shown by the peak in the standard-deviation of $r$ in the inset of Figure~\ref{fig::r_x_den_221}. The peak increases with the network size. For small values of $\rho$ synchronization is maintained, $r>0$, as conformists synchronize their phases with each other and the small fraction of contrarians dephase from their neighbors without destroying global synchronization. In this situation, the average phase of contrarians and conformists create a periodic alternating wave over time, an interesting mechanism that resembles, for instance, the oscillation of populations of predators and preys which characterizes the classical Lotka-Volterra model \cite{Castillo01}. Figure~\ref{fig::phi_x_t_nets_over_time_pairwise} shows an example where contrarians (the central layer of the networks in the upper part) are in opposition to their first neighbors conformists, which in turn try to synchronize with them. The sequence of networks in the figure are snapshots of the oscillators and their phase over time. Before all conformists could match their phases with contrarians, the latter already have an opposite phase. The periodic wave that conformists and contrarians create is clear in the lower part of Figure~\ref{fig::phi_x_t_nets_over_time_pairwise} which shows the average phase of different types of oscillators.

As the fraction of contrarians overcomes a certain threshold, the effect of contrarians spreads over the entire network completely suppressing global synchronization. This suppression is a consequence of an increasing fraction of contrarian/contrarian interactions, which naturally tend to be dephased, reinforcing their impact. For $\rho>0.15$, neither synchronization ($r=0$) nor a periodic wave is observed (see Supplementary Fig.~S2). 

\subsection{Contrarians as hubs}

The phenomenon of synchronization is known to result from the interplay between the network topology and the dynamics of oscillators \cite{Gardenes11,Assenza11}. In this section, we discuss the improvement in the desynchronization efficiency by distributing the contrarians among the oscillators (nodes) with higher degree and compare this strategy with the random distribution case discussed above.

We start considering the case of a random graph (Erd\H{o}s-R\'{e}nyi (ER) network), characterized by a Poisson distribution of degree. As shown in the Supplementary Fig.~S3, in spite of the narrow degree distribution, the fraction of contrarians necessary to reduce synchronization is reduced to one third ($\rho=0.05$) when contrarians sit at the nodes with higher degree. In this case, the disturbing effect of contrarians occurs even for smaller fractions $\rho$ and synchronization is efficiently suppressed. In scale-free networks, where degree distribution is scale free, and highly connected nodes are more frequent, the assignment of hubs as contrarians favors desynchronization (see Supplementary Fig.~S4).

\begin{figure*}[t]
\includegraphics[width=2\columnwidth]{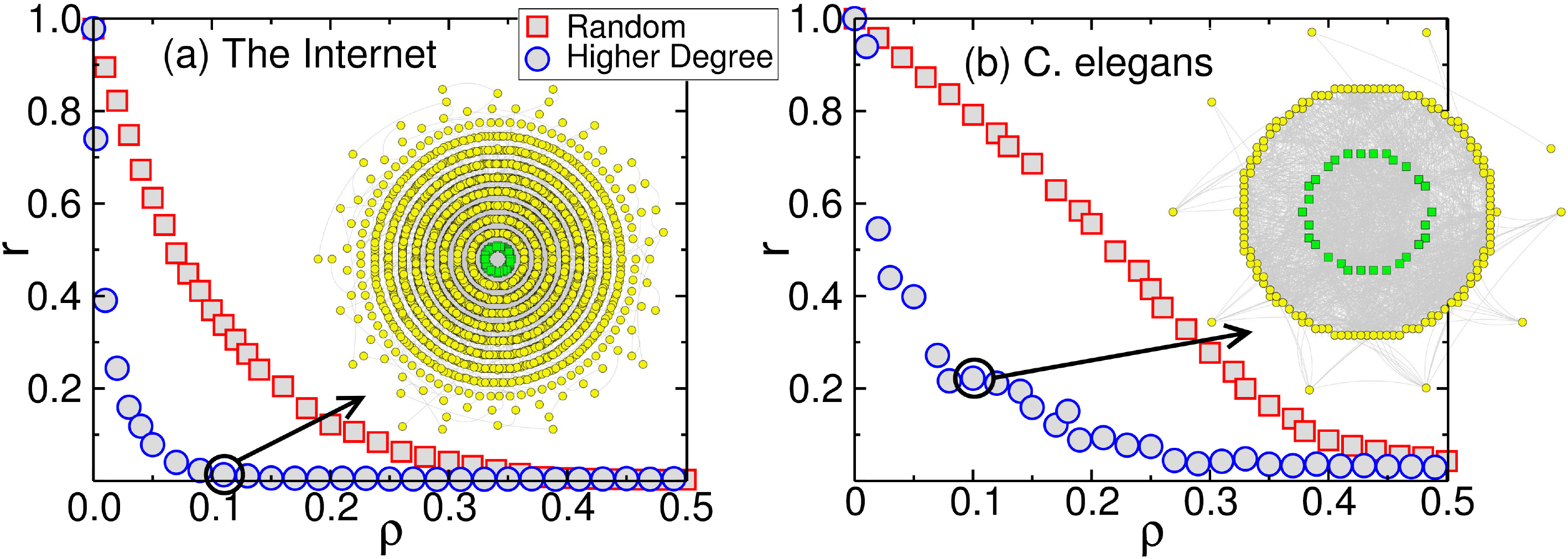}
 \caption{\textbf{Pairwise contrarians on real networks.} Fraction of contrarian oscillators randomly assigned and based on the degree for different networks: (a) the Routers of the Internet, (b) neurons on \emph{C. elegans}. The insets are snapshots of the referred networks where each concentric layer $i$, from inside to outside, contains the $i$th neighbors of the contrarians. The color of each node represents its type: contrarians (green) and conformists (yellow). A coupling strength of $\lambda=4.0$ has been used.}
\label{fig::r_x_den_221asst_neuron_internet}
\end{figure*}

In real networks, the presence of communities and other features, such as assortativity and clustering, also play a role in synchronization~\cite{Oh05,Arenas06}.
Here we consider two real networks and show that the same behavior holds (see Figure~\ref{fig::r_x_den_221asst_neuron_internet}). The first one is the network of Routers in the Internet. This network is believed to have grown through the mechanism of preferential attachment, being characterized by a scale-free degree distribution of exponent $\gamma=3.00$~\cite{Faloutsos99}. Moreover, it has been shown to have a hub dominant structure, where many hubs share low degree neighbors~\cite{Lancichinetti10}. In this case, similarly to scale-free networks, hubs play a major role in the synchronization, and only 10\% of them are necessary to suppress synchronization. Highly connected contrarians have a disordering effect on a great number of conformists and a few hubs control the phase of the entire set of oscillators. The second one is the neuronal network of the organism \emph{C. elegans}, a Small-World network and the largest network of neurons that has been totally mapped~\cite{Watts03}. There, random or degree-based distribution of contrarians suppress synchronization, although the network seems to be more resilient to such control than ER networks. The presence of functional communities of neurons that are highly connected within themselves might be the cause of this resistance~\cite{Hu10}. It is interesting to note that this biological system, evolved under evolutionary pressure, has converged to a resilient structure regarding synchronization. 

\section{Discussion}
The use of agents to control the dynamics of a network has recently been a subject of research interest~\cite{Liu11,Yan12}. In this work, we have shown that global synchronization can be suppressed with local agents (contrarians) which systematically dephase from their nearest neighbors. We show that solely local information is required to efficiently avoid a coherent oscillatory state. If instead, global information is considered, the set of oscillators splits into two oscillatory states with different phases and a global coherent state is still possible. Analyzing the impact of the network topology in desynchronization we concluded that, when contrarians sit at the nodes of higher degree, the process is more efficient than in the random case. Even with random graphs, characterized by a narrow degree distribution, the degree strategy reduces to one third the amount of required contrarians to suppress synchronization. 
We also show that the synchronization of real networks that present underlying features such as communities and dominant hubs is also suppressed with the use of contrarians.

The present work is the first attempt to understand the interplay between the desynchronization dynamics and the topology when considered to mimic real systems. The Kuramoto model provides a standard framework to study synchronization, however it entails several approximations when discussing real networks. For example, social systems are composed of adaptive agents that might change their strategy over time and avoid being trapped in a locked state. Developments build up on top of this work should account for further details on the coupling scheme and on the contrarian dynamics. For instance, here we have focused on the distribution of contrarians, but recent works have shown that rewiring interventions, such as swapping, adding, or removing edges, have a crucial role in the collective dynamics~\cite{Volman10,Volman11}. Nevertheless, general conclusions can be drawn shedding light on the problems discussed in the introduction. For instance, as referred, some Routers on the Internet have a special algorithm implemented to avoid synchronization. We have shown that placing contrarian Routers as hubs on the network could optimize the fraction of proactive Routers necessary to prevent global synchronization. Also, modifying the coupling mechanism between two Routers and the contrarian strategy implemented, it becomes possible to extend our work to define the best location of contrarians on the Internet.

Regarding the development of brain pacemakers, our study suggests that a set of small size devices spread throughout the brain, and solely tracking the phase of neighboring neurons, would be more effective to prevent a seizure than monitoring the global state. After the mapping of a neural network \cite{Eguiluz05}, and the characterization of the coupling dynamics between neurons, our work also gives helpful hints about the minimum amount and the optimal spatial distribution, of these active devices.

Interesting applications could also be found in social dynamics. Whenever a ``social synchronization'' is achieved, such as clapping after a speech, a small amount of influential agents can be trained to prevent this synchronization. Political opponents could be spread in the crowd to avoid a proper salutation simply by ``dephasing'' their claps with their close neighbors. The same method might be used to prevent a synchronized walk on a bridge where instructed actors could walk dephased from others. Evidently crowd behavior control is a very hard task \cite{Lorenz11}, but here we show that it could theoretically be achieved.

\section{Methods}
Equations \ref{eq::kuramoto}, \ref{eq::meanfield}, and \ref{eq::pairwise} have been numerically solved using a fourth order Runge-Kutta method with discrete time steps $\delta t=0.001$. The complex order parameter was computed in the interval $t\in\left[90,100\right]$ using the average value of Equation~\ref{eq::order.param}. For all considered cases, natural frequencies of oscillators have been uniformly distributed between $-0.5$ and $0.5$ and initial phases have also been uniformly distributed between $-\pi$ and $\pi$. A coupling strength of $\lambda=2.0$ has been used. For the Internet and the \emph{C. elegans}, natural frequencies and initial phases have been distributed uniformly between $-0.1$ and $0.1$, and between $-\frac{\pi}{2}$ and $\frac{\pi}{2}$, respectively, and a coupling strength of $\lambda=4.0$ has been used. 

\begin{table}[h]
\resizebox{\columnwidth}{!}{%
\begin{tabular}{lccc}
\toprule
Network & Nodes & Average Degree &  Max Degree \\ 
\hline
The Internet & $40028$ & $2.36$ & $259$ \\
\emph{C. elegans} & $283$ & $17.39$ & $115$\\
\hline
\bottomrule
\end{tabular}}
\caption{Number of nodes, average degree, and maximum degree for network of Routers in the Internet and the neural network of the \emph{C. elegans}.}
 \label{tab::measures}
\end{table}

Networks of oscillators have been constructed as undirected ER networks of average degree four, unless otherwise stated. The network of Internet Routers has been analyzed through data retrieved from the Opte Project that represents all the communication among $40028$ Routers on January 15th of 2005~\cite{Lyon03}. Each node of this network is a Router with an associated IP address and the links (edges) are established between two IP address which have communicated at least once. The network of neurons was constructed through data obtained on the WormWeb website \cite{Bhatla09} and is mostly based on the work by Chen \emph{et al} \cite{Chen06}. In this network, links have been established whenever an interaction between neurons has been registered, regardless of their type or direction. Other measures regarding these networks are available in Table~\ref{tab::measures}.

All results have been averaged over several samples. The error bars were omitted in all figures, being smaller than the symbols. For the Internet and the \emph{C. elegans} only the natural frequencies and initial phases change among samples.  
Plots in Figure~\ref{fig::r_x_l_meanfield_pairwise} are constructed using the average value over $100$ networks of size $\mbox{N = 1000}$. Figure~\ref{fig::r_x_den_221} shows averages of $5000$ networks of size $\mbox{N = 1000}$, $1000$ networks of size $N = 5000$, and $600$ networks of size $N = 10000$. In Figure~\ref{fig::phi_x_t_nets_over_time_pairwise}, we represent a single network of $N=200$ with $\rho=0.1$ and $\lambda=1.0$. For Figure~\ref{fig::r_x_den_221asst_neuron_internet}, plot~(a) is an average over $200$ initial distributions of phases and frequencies, and plot~(b) is an average over $10$.

\vspace{10.0pt}
\textbf{Acknowledgments.} Authors would like to thank the Swiss National Science Foundation under contract 200021 126853, the CNPq, Conselho Nacional de Desenvolvimento Científico e Tecnológico - Brasil, the CNPq/FUNCAP Pronex grant, and the INCT-SC for financial support, and the Cuttlefish Project, responsible for the network visualization tool.

\clearpage

\begin{minipage}{2\linewidth}
\begin{center}
\Large{\textbf{How to suppress undesired synchronization}}
\vspace{5.0pt}

\large{Supplementary Information}
\end{center}

\end{minipage}
\renewcommand{\thefigure}{S\arabic{figure}}
\renewcommand{\thetable}{S1}

\pagenumbering{gobble}
\setcounter{figure}{0}

\begin{figure}[h]
\includegraphics[width=\columnwidth]{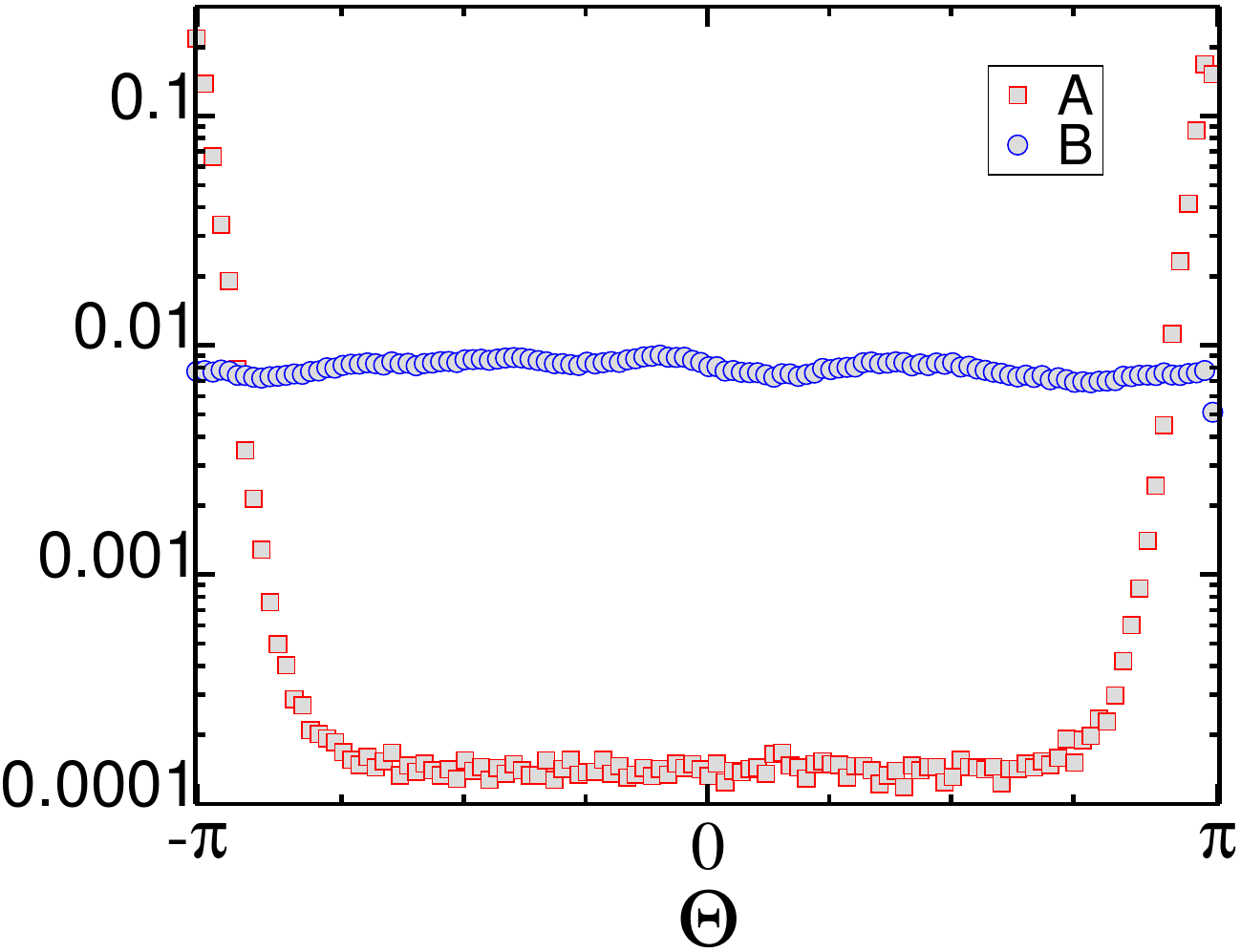}
\caption{\textbf{Distribution of oscillator phases}. Histogram of the phase of oscillators for $t=100$. Curves represent oscillators according to the mean-field (A) and pairwise (B) models. While mean-field contrarians are frozen in $-\pi$ and $\pi$, pairwise contrarians are uniformly distributed through all phases. Results are averages over 1000 ER networks of 1000 oscillators each, where $\rho=0.1$ and $\lambda=1.0$.}
\label{fig::hist_phase_121_221}
\end{figure}

\begin{figure*}[h]
\includegraphics[width=2\columnwidth]{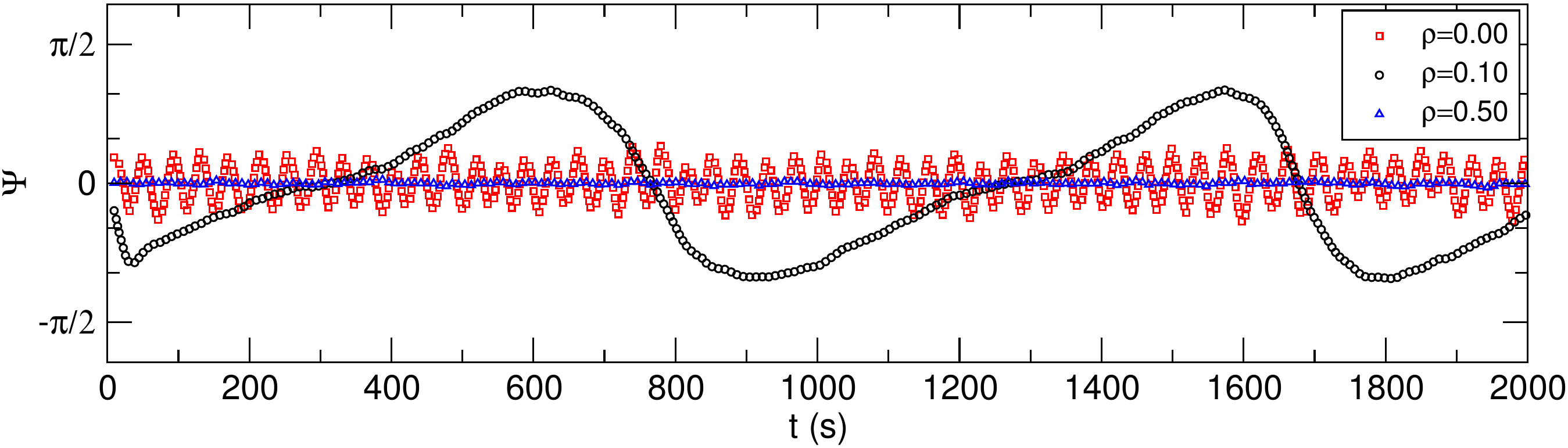} %
\caption{\textbf{Average phase of oscillators with different fractions of pairwise $\rho$ contrarians}. Time dependence of the average phase of oscillators for different fractions of randomly assigned contrarians. The amplitude of the wave goes to zero with the fraction contrarians oscillators.}

\label{fig::phi_x_t}
\end{figure*}

\begin{figure}
\includegraphics[width=\columnwidth]{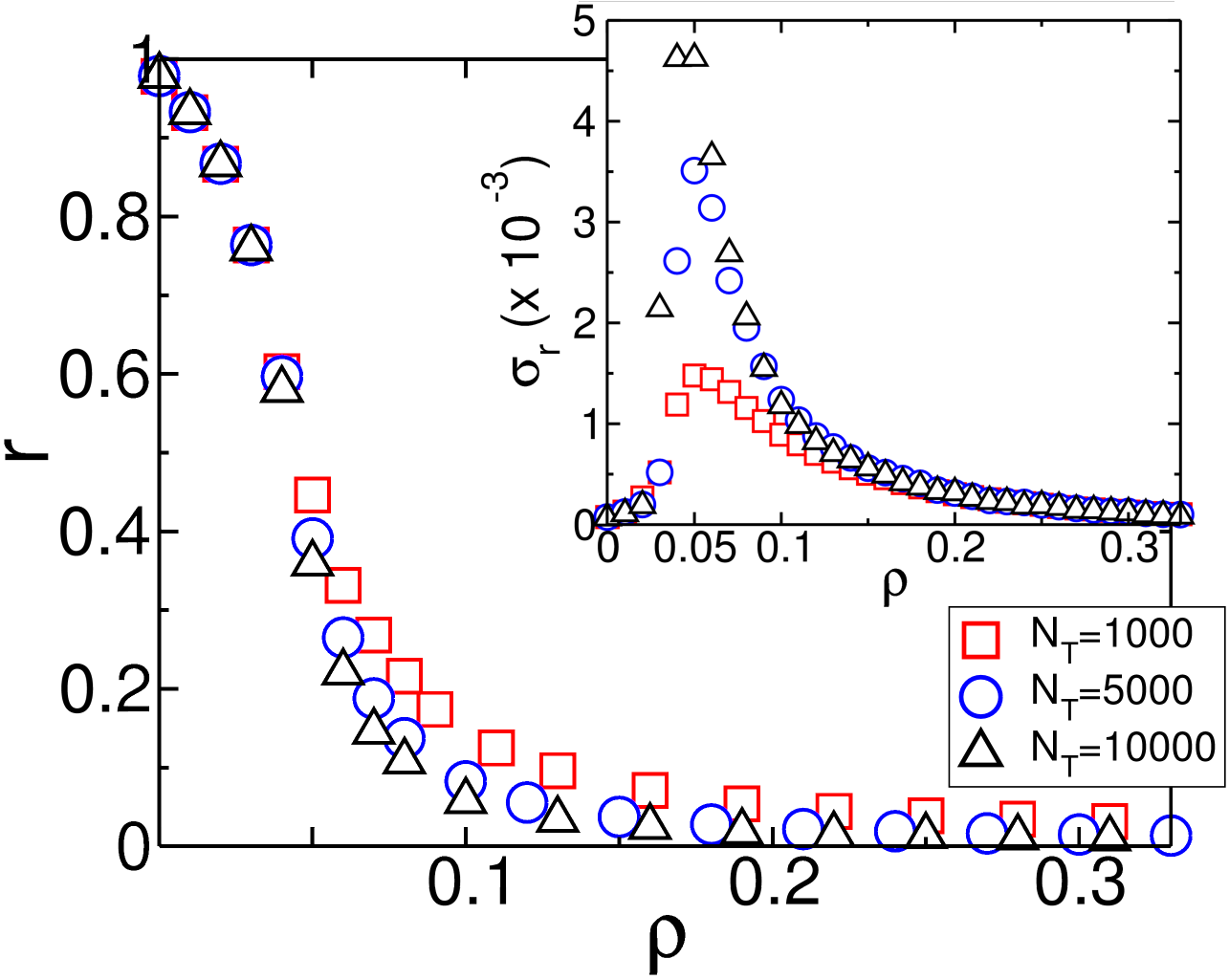}
\caption{\textbf{Impact of pairwise contrarians assigned to the nodes of highest degrees on the synchronization of ER networks of average degree four.} Order parameter $r$ dependence on the fraction of contrarians $\rho$ for different network sizes showing a suppression of synchronization after the introduction of pairwise contrarians. The inset is the standard-deviation of $r$ showing a transition around $\rho=0.05$, much smaller than ramdomly assigned contrarians.}
\label{fig::r_x_den_221_asst}
\end{figure}

\begin{figure*}[h]
\includegraphics[width=2\columnwidth]{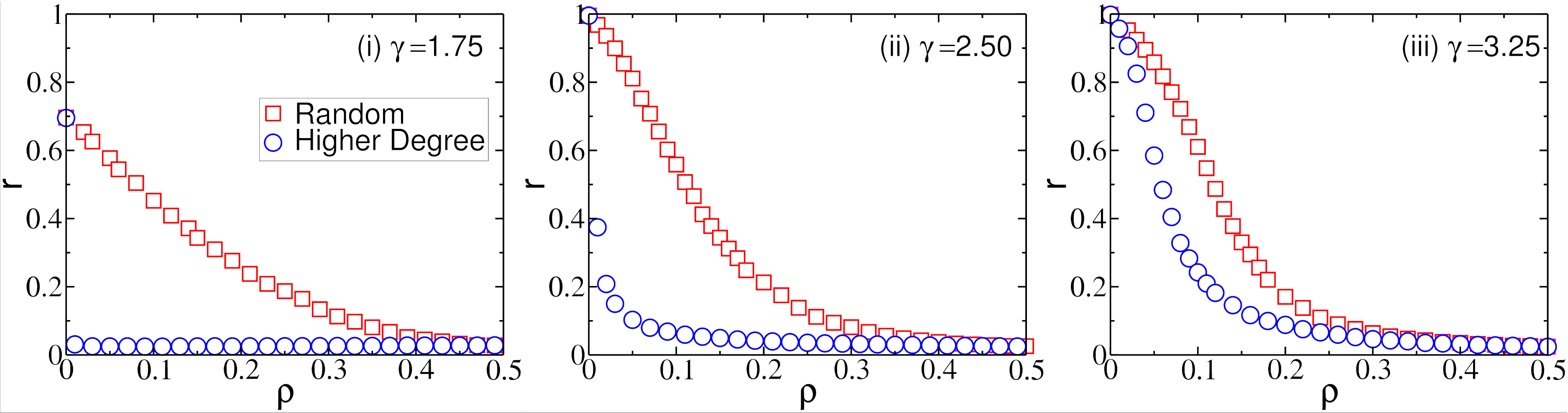} %
\caption{\textbf{Scale-free networks with contrarians}. Fraction of pairwise contrarian oscillators assigned randomly and based on their degree to scale-free networks of different degree exponent $\gamma$, namely, $1.75$, $2.5$, and $3.25$.}
\label{fig::r_x_den_gammas}
\end{figure*}

\end{document}